\documentclass[reprint,aps,prc,twocolumn,superscriptaddress,floatfix,amsmath,amssymb,10pt]{revtex4-2}
\DeclareMathAlphabet{\mathpzc}{OT1}{pzc}{m}{it}
\usepackage{bm}
\usepackage{dcolumn}
\usepackage{amsthm}
\usepackage{amsmath}
\usepackage{amssymb}
\usepackage{graphicx}
\usepackage{xcolor}
\usepackage{fix-cm}
\usepackage{mathptmx} 
\usepackage[T1]{fontenc}
\usepackage[colorlinks,allcolors=blue]{hyperref}
\makeatletter
\def\NAT@def@citea{\def\@citea{\NAT@separator}}
\makeatother

\begin{document}

\title{Pauli energy contribution to nucleus-nucleus interaction}

\author{A.S. Umar}\email{umar@compsci.cas.vanderbilt.edu}
\affiliation{Department of Physics and Astronomy, Vanderbilt University, Nashville, Tennessee 37235, USA}
\author{C. Simenel}\email{cedric.simenel@anu.edu.au}
\affiliation{Department of Theoretical Physics and Department of Nuclear Physics, Research School of Physics,
The Australian National University, Canberra ACT 2601, Australia}
\author{K. Godbey}\email{godbey@frib.msu.edu}
\affiliation{Cyclotron Institute, Texas A\&M University, College Station, TX 77843, USA}
\altaffiliation{Current address: Facility for Rare Isotope Beams, Michigan State University, East Lansing, Michigan 48824, USA }
\date{\today}


\begin{abstract}
\edef\oldrightskip{\the\rightskip}

\begin{description}
	\rightskip\oldrightskip\relax
	\setlength{\parskip}{0pt} 
	\item[Background]
	The Pauli exclusion principle plays a crucial role as a building block of many-body quantal systems
	comprised of fermions. It also induces a ``Pauli repulsion'' in the interaction
	between di-nuclear systems. It has been shown in [Phys. Rev. C{\bf{95}}, 031601 (2017)]  that the Pauli
	repulsion widens the nucleus-nucleus potential barrier, thus hindering sub-barrier fusion.
	\item[Purpose]
	To investigate the proton and neutron contributions to the Pauli repulsion, both in the bare potential neglecting shape polarization and transfer between the reactants, as well as in the dynamical potential obtained by accounting for such dynamical rearrangements. 
	\item[Methods]
	As the basis of our study we utilize the Pauli kinetic energy (PKE) obtained by studying
	the nuclear localization function (NLF). Recently this approach has been generalized to
	incorporate all of the dynamical and time-odd terms present in the nuclear energy density
	functional. This approach is employed in the density
	constrained Hartree-Fock (DCFHF) and in the density constrained time-dependent Hartree-Fock
	(DC-TDHF) microscopic methods.
	\item[Results]
	The PKE spatial distribution shows that a repulsion occurs in the neck between the nuclei when they first touch. Inside the barrier, neutrons can contribute significantly more to the Pauli repulsion in neutron-rich systems. Dynamical effects tend to lower the Pauli repulsion near the barrier. Proton and neutron dynamical contributions to the PKE significantly differ inside the barrier for asymmetric collisions, which is interpreted as an effect of multinucleon transfer.
	\item[Conclusions]
	The PKE is shown to make a significant contribution to nuclear interaction potentials.
	Protons and neutrons can play very different roles in both the bare potential and in the dynamical rearrangement. Further microscopic studies are required to better understand the role of transfer and to investigate the effect of pairing and deformation. 
\end{description}
\end{abstract}
\maketitle


\section{Introduction}

One of the most important features of quantum mechanics is the classification of
particles based on their spin content as bosons and fermions. For quantum
systems composed of fermions the Pauli exclusion principle states that no
two identical particles can occupy the same state. This principle, which was
introduced by Pauli for electrons~\cite{pauli1925}, was subsequently generalized to all
fermions via the spin-statistics theorem~\cite{fierz1939,pauli1940}. For complex many-body systems
the Pauli principle manifests itself as a requirement that the wave function
representing the system be antisymmetric under exchange of particles.
In density 
functional theory (DFT) calculations for electronic systems, the construction of the so-called orbital-free DFT functionals contains a Pauli kinetic energy term in the energy decomposition. 
The orbital based determination of the Pauli kinetic energy is found to be essential for gauging the
accuracy of these empirical Pauli functionals~\cite{liu2019b,patra2019}. Similarly,
the identification of the contribution of Pauli repulsion in DFT calculations of
intermolecular interaction energies is of current interest in 
chemistry~\cite{wu2009,raupach2015,sarsa2019,andrada2020,horn2016}.

Atomic nuclei, composed of neutrons and protons, must also adhere to the Pauli exclusion principle.
This is strictly implemented in microscopic calculations of nuclear structure, primarily
by adopting determinental many-body wavefunctions. 
Similar to intermolecular interactions, one expects, in the case of collisions of atomic nuclei,
a Pauli repulsion induced by the Pauli exclusion principle 
between nucleons of different nuclei.
Although this Pauli repulsion is often neglected in calculations of nucleus-nucleus potentials based on the observation that the overlap between collision partners at distances close to the fusion barrier radius is usually small, it has been shown that the Pauli repulsion can play a significant role in systems with large $Z_1Z_2$ (as more overlap is required to compensate the larger Coulomb repulsion), as well as at energies well below the Coulomb barrier~\cite{simenel2017}. Indeed, in the latter case, fusion occurs via tunneling through a barrier widened by Pauli repulsion as the inner turning point corresponds to a substantial nuclear overlap with a more compact 
shape~\cite{dasso2003,umar2012a}.
As a result, Pauli repulsion provides a natural, though partial, explanation to experimentally observed fusion hindrance at deep sub-barrier energies~\cite{jiang2002,dasgupta2007,stefanini2010} (see~\cite{back2014} for a review). Note also that collisions well above the barrier are expected to be sensitive to Pauli repulsion as well~\cite{fliessbach1971,brink1975,beck1978,sinha1979}.

Several techniques have been used in the past, with various success, to 
incorporate the Pauli exclusion principle for the calculation of ion-ion
potentials. 
These include techniques to orthogonalize overlapping nucleon wave-functions~\cite{fliessbach1971,brink1975},
though the orthogonalization itself usually induces a spurious change of the density and thus of the nuclear attraction~\cite{fliessbach1975,simenel2017}.
One approach is to replace the kinetic energy density using expressions based on the Thomas-Fermi 
 approximation~\cite{zint1975,brink1978,denisov2010}.
 In this approach, however, the effect of other momentum dependent terms as well as the spin-orbit interaction of the
effective nucleon-nucleon interaction on the Pauli repulsion are not accounted for.
A folding density method for $\alpha$-clusters has also been developed recently~\cite{cheng2019,cheng2020}.
Attempts were made with the resonating group method~\cite{aoki1983,wada1987,tohsaki1975,tang1978} which, in principle, provides a theoretical approach to construct internuclear potentials with full antisymmetrization. However, such calculations have thus far been limited to light systems and direct reactions due to their complexity.


A natural way to describe heavy ion collisions, while accounting exactly for the Pauli exclusion principle, is to use microscopic theories in which the wave-function of the entire system is a Slater determinant of nucleon wave-functions (mean-field approximation) or a sum of such determinants (beyond mean-field approaches). 
In particular, the time-dependent Hartree-Fock theory, that describes the evolution of each particle in the self-consistent mean-field produced by all particles, is appealing as it does not rely on a nucleus-nucleus potential. 
The validity of the mean-field approximation is based on the expectation that the Pauli principle plays an important role in
simultaneously building up a time-dependent mean-field and suppressing the propagation of the
strong nucleon-nucleon interaction terms. 
Although the TDHF theory has been successful in describing various reaction mechanisms (see~\cite{simenel2012,simenel2018,stevenson2019,sekizawa2019} for recent reviews), its main drawback is that it does not account for tunneling of the many-body wave-function. Though imaginary-time mean-field methods are expected to overcome this limitation~\cite{levit1980c,reinhardt1981a}, only applications to simple systems have been achieved so far~\cite{arve1987,mcglynn2020}.
Therefore, descriptions of sub-barrier fusion nowadays are still based on calculations of barrier penetration with nucleus-nucleus potentials (see~\cite{hagino2012,back2014} for reviews).

It is  possible to compute nucleus-nucleus potentials microscopically, using a single Slater determinant to describe the many-body wave-function and account for Pauli repulsion. 
As an example, one can use the two-center shell model~\cite{nesterov2013,nesterov2018} or the constrained Hartree-Fock method~\cite{skalski2007}. However, both methods assume an adiabatic approximation. The latter, which is expected to hold in nuclear fission as it is a slow process, is questionable in the context of ``faster'' reactions such as fusion.
For this reason, nucleus-nucleus potentials are more commonly computed with the frozen density approximation in which the nuclei are assumed to keep their ground-state densities. 
Fusion barriers can be computed with microscopically obtained nuclear densities by evaluating the
interaction energy between the nuclei at an internuclear separation, $R$, while keeping
these densities frozen~\cite{brueckner1968}. Subsequently, the dynamics can be included via the coupled-channels
approach~\cite{hagino1999}.
Using this approximation with HF ground-state densities of the collision partners, one gets the frozen Hartree-Fock (FHF) method~\cite{denisov2002,skalski2007,washiyama2008,simenel2008,guo2012,vophuoc2016} which, completed with coupled-channel calculations to account for rearrangement of nuclear densities, provides a reasonable description of fusion at near-barrier energies~\cite{simenel2013b}. 
FHF potentials, however, do not account for the Pauli repulsion as, at small $R$, it involves two overlapping Slater determinants (one per HF ground-state). Thus FHF potentials should not be used to describe fusion well below the barrier, or in systems with large $Z_1Z_2$~\cite{simenel2017}.
For this reason, we recently developed the  so-called density-constrained frozen Hartree-Fock (DCFHF)
method which is based on the frozen density approximation, while describing the many-body system with a single Slater determinant, thus accounting for the Pauli exclusion principle exactly~\cite{simenel2017}. DCFHF is the static counterpart of the density constrained time-dependent Hartree-Fock (DC-TDHF) method which is used to determine nucleus-nucleus potentials including dynamical effects~\cite{umar2006b}. A comparison between FHF (static without Pauli repulsion), DCFHF (static with Pauli repulsion) and DC-TDHF (dynamics with Pauli repulsion) potentials thus provides valuable information on the respective roles of Pauli repulsion and dynamical density rearrangement. 
Nevertheless, the information one can access with this method is only global as we are dealing with quantities that are integrated over space. In order to get a deeper insight into the microscopic origin of Pauli repulsion, it is therefore desirable to express it in terms of a local quantity. 

In this work, we achieve this using the nucleon localization function (NLF) which  allows for a more precise characterization of the Pauli energy, and in particular its spacial distribution. 
The electronic localization function (ELF) was originally developed in condensed matter physics~\cite{becke1990,burnus2005,savin2005}
as a way to visualize the localized electronic groups in atomic and molecular systems~\cite{jerabek2018}. The same idea was
introduced to nuclear physics to bring out the clustering features of the nuclear densities obtained in 
mean-field calculations~\cite{reinhard2011} and employed to bring out the cluster features of light 
nuclei~\cite{reinhard2011,ebran2017}, study of reactions~\cite{schuetrumpf2017}, and shell effects in 
fission~\cite{zhang2016,scamps2018,scamps2019,matheson2019,sadhukhan2020}.
In a recent work, the NLF function was generalized to include all of the time-odd and spin dependent terms of the
Skyrme energy density functional~\cite{li2020}. As a byproduct of these developments we also have an accurate
way to calculate the Pauli energy in a similar way it is employed in DFT
calculations~\cite{liu2019b,patra2019} for electronic systems
 while including all the nuclear energy density functional contributions.

We use the NLF approach to demonstrate the contribution of Pauli energy in nucleus-nucleus potentials calculated with the DCFHF and DC-TDHF methods. 
The NLF is expressed in terms of a  Pauli localization function (PLF)
to  visualize these effects. 
In Sec.~\ref{sec1} we give a brief outline of the methods employed to compute the Pauli contribution to the ion-ion interaction barriers and we describe the Pauli energy calculations using the
NLF. This is followed by the results of these calculations in Sec.~\ref{sec2}. Conclusions are summarized in Sec.~\ref{sec3}.

\section{Method}\label{sec1}
In this section we briefly outline the formalisms and methods used in our calculations. Further details can be found
in the cited references.

\subsection{Microscopic methods}
In order to investigate the Pauli energy in heavy-ion fusion reactions,
we employ microscopic methods to compute the interaction between nuclei.
Following the idea of Brueckner \textit{et al.}~\cite{brueckner1968}, 
the bare nucleus-nucleus potential (i.e., without dynamical rearrangement) is computed from an energy density functional (EDF)  $E[\rho]$
written as a space integral of an energy density $\mathpzc{H}[\rho(\mathbf{r})]$
\begin{equation}
E[\rho]=\int d\mathbf{r}\; \mathpzc{H}\left[\rho(\mathbf{r})\right]\,.
\end{equation}
The bare potential is obtained by requiring frozen ground-state densities $\rho_{i}$ 
of each nucleus ($i=1,2$) which we compute
using the Hartree-Fock (HF) mean-field approximation. 
One advantage of this method is that it does not introduce new parameters other than those of the EDF.
Indeed, the same Skyrme EDF~\cite{skyrme1956} is used both in HF calculations and to compute the bare potential.

Neglecting the Pauli exclusion principle between nucleons in different nuclei
leads to the usual FHF 
potential~\cite{denisov2002,skalski2007,washiyama2008,simenel2008,guo2012,vophuoc2016}
\begin{equation}
V_\mathrm{FHF}(\mathbf{R})=\int d\mathbf{r}\;\mathpzc{H}\left[\rho_1(\mathbf{r})+\rho_2(\mathbf{r}-\mathbf{R})\right]- E[\rho_1] -E[\rho_2]\,,
\label{eq:frozen}
\end{equation}
where $\mathbf{R}$ is the distance vector between the centers of mass of the nuclei.
To account for the Pauli repulsion in the bare potential, we use instead the DCFHF method~\cite{simenel2017}.
The Pauli exclusion principle is included exactly by allowing the single-particle states, 
comprising the combined nuclear density, to reorganize
to attain their minimum energy configuration and be properly antisymmetrized as the many-body
state is a Slater determinant of all the occupied single-particle wave-functions.
The HF minimization of the combined system is thus performed subject to the constraint that the
local proton ($p$) and neutron ($n$) densities do not change:
\begin{equation}
\delta \left\langle \ H - \sum_{q=p,n}\int\, d\mathbf{r} \ \lambda_q(\mathbf{r})  \left[\rho_{1_q}(\mathbf{r})+\rho_{2_q}(\mathbf{r}-\mathbf{R})\right] \ \right\rangle = 0\,,
\label{eq:var_dens}
\end{equation}
where the $\lambda_{n,p}(\mathbf{r})$ are Lagrange parameters at each point 
of space constraining the neutron and proton local densities.
This equation determines a unique Slater determinant $|\Phi(\mathbf{R})\rangle$.
Assuming the potential to be central leads to the expression
\begin{equation}
V_{\mathrm{DCFHF}}(R)=\langle\Phi(\mathbf{R}) | H | \Phi(\mathbf{R}) \rangle-E[\rho_1]-E[\rho_2]\,.
\label{eq:vr}
\end{equation}

The resulting DCFHF bare potentials thus account for the Pauli repulsion, widening the fusion barrier and producing a potential pocket at short distance $R$ which is not present in  FHF potentials~\cite{simenel2017}. 
Consequently, comparisons between FHF and DCFHF bare potentials allow us to study the effects of the Pauli principle for frozen nuclear densities. 
These potentials, however, do not account for any dynamical rearrangement of the densities induced. 

TDHF calculations, on the other hand, account for such rearrangement, at the mean-field level, in particular those produced by couplings to vibrational~\cite{flocard1981,simenel2013a,simenel2013b} and rotational modes~\cite{simenel2004,umar2006d}, as well as nucleon transfer through the neck~\cite{simenel2008,umar2008a,simenel2010,sekizawa2013,vophuoc2016,godbey2017,jiang2020}.
Nucleus-nucleus potentials extracted from TDHF calculations~\cite{umar2006b,washiyama2008,guo2012,washiyama2020} then account for both dynamical effects as well as the Pauli exclusion principle.
In order to further investigate the full dynamics we have thus used the
DC-TDHF method, where the densities are taken directly from the TDHF evolution of the system
 and the same constraint procedure used in DCFHF is employed (see~\cite{umar2006b} for details):
\begin{eqnarray}
E_{\mathrm{DC-TDHF}}(t)&=&\underset{\rho}{\min}\left\{E[\rho_n,\rho_p]+\right.\nonumber\\
&&\sum_{q=p,n}\int\, d\mathbf{r} \ \lambda_q(\mathbf{r})  \left.\left[\rho_{q}(\mathbf{r})-\rho_{q}^{\mathrm{TDHF}}(\mathbf{r},t)\right]\right\}\,.\nonumber
\end{eqnarray}
The same TDHF evolution gives access to the time evolution of the distance $R(t)$ between the centers of mass of the fragments. 
The potential is then obtained as in Eq.~(\ref{eq:frozen}) by removing the binding energy of the HF ground-states,
\begin{equation}
V_{\mathrm{DC-TDHF}}(R)=E_{\mathrm{DC-TDHF}}(R)-E[\rho_1]-E[\rho_2]\,.
\label{eq:vrdctdhf}
\end{equation}

The DC-TDHF approach accounts for microscopic effects associated with the Pauli exclusion principle, such as the splitting of orbitals with some
states contributing attractively (bounding) and some repulsively (antibounding) to the potential~\cite{umar2012b}.
It should be noted also that, due to their dynamical nature, density rearrangements naturally depends on the energy of the collision, inducing an energy dependence to the potential~\cite{washiyama2008,umar2009a,keser2012,umar2014a}. One limitation of  methods relying on TDHF evolutions is that the latter do not account for many-body tunneling. Thus, the energy dependence of the potential at sub-barrier energies is unknown. Nevertheless near and sub-barrier fusion cross-sections have been computed using DC-TDHF potentials determined from TDHF evolutions at near-barrier central collisions, showing overall good agreement with experiment~\cite{umar2009b,oberacker2013,simenel2013a,jiang2014,umar2014a,guo2018,guo2018b,scamps2019b,godbey2019b}.

\subsection{The nucleon localization function (NLF)}
The measure of localization has been originally developed in the context of
a mean-field description for electronic systems~\cite{becke1990,burnus2005,savin2005},
and subsequently introduced to nuclear systems~\cite{reinhard2011,li2020}.
We first realize that a fermionic mean-field state is fully characterized by the one-body density-matrix
$\rho_q(\mathbf{r}s,\mathbf{r'}s')$.
The probability of finding two nucleons with the same spin at spatial
locations $\mathbf{r}$ and $\mathbf{r'}$ (same-spin pair probability) for
isospin $q$ is proportional to
\begin{equation}
	P_{qs}(\mathbf{r},\mathbf{r}') =
	\rho_q(\mathbf{r}s, \mathbf{r}s)\rho_q(\mathbf{r}'s, \mathbf{r}'s)
	-
	|\rho_q(\mathbf{r}s,\mathbf{r}'s)|^2\,,
\end{equation}
which vanishes for $\mathbf{r}=\mathbf{r'}$ due to the Pauli exclusion principle.
The conditional probability for finding a nucleon at
$\mathbf{r'}$ when we know with certainty that another nucleon with the same
spin and isospin is at $\mathbf{r}$ is proportional to
\begin{equation}
	\label{conditionalProb}
	R_{qs}(\mathbf{r},\mathbf{r}')
	=
	\frac{P_{qs}(\mathbf{r},\mathbf{r}')}{\rho_q(\mathbf{r}s,\mathbf{r}s)}\,.
\end{equation}
The short-range behavior of $R_{qs}$ can be 
obtained using techniques similar to the local density approximation~\cite{reinhard2011,li2020}. The
leading term in the expansion yields the localization measure
\begin{equation}\label{eq:prob_D}
    D_{qs_{\mu}} = \tau_{qs_{\mu}}-\frac{1}{4}\frac{\left|\boldsymbol{\nabla}\rho_{qs_{\mu}}\right|^2}{\rho_{qs_{\mu}}}
    -\frac{\left|\mathbf{j}_{qs_{\mu}}\right|^2}{\rho_{qs_{\mu}}}\,.
\end{equation}
This measure is the most general form that is appropriate for deformed nuclei and without assuming
time-reversal invariance, thus also including the time-odd terms important in applications such as
cranking or TDHF. The densities and currents are given in their most unrestricted  form~\cite{engel1975,bender2003,perlinska2004}
for $\mu$-axis denoting the spin-quantization axis by~\cite{li2020}
\begin{subequations}\label{eq:density_relation}
	\begin{align}
		\rho_{q s_\mu}(\mathbf{r}) &= \frac{1}{2}\rho_q(\mathbf{r}) + \frac{1}{2}\sigma_\mu {s}_{q\mu}(\mathbf{r})\,, \\
		\tau_{q s_\mu}(\mathbf{r}) &= \frac{1}{2}\tau_q(\mathbf{r}) + \frac{1}{2}\sigma_\mu {T}_{q\mu}(\mathbf{r})\,, \\
		\mathbf{j}_{q s_\mu}(\mathbf{r}) &= \frac{1}{2}\mathbf{j}_q(\mathbf{r}) + \frac{1}{2}\sigma_\mu \mathbb{J}_q(\mathbf{r})\cdot \mathbf{e}_\mu\,,
	\end{align}
\end{subequations}
where $\sigma_\mu =2s_\mu=\pm1$ and $\mathbf{e}_\mu$ is the unit vector in the direction of the $\mu$-axis.
The explicit expressions of the local densities and currents are given in Refs.~\cite{engel1975,li2020}.
We note that the localization measure includes the spin-density ${s}_{q\mu}(\mathbf{r})$, the time-odd part of the kinetic density ${T}_{q\mu}(\mathbf{r})$,
as well as the full spin-orbit tensor $\mathbb{J}_q(\mathbf{r})$, which is a pseudotensor. In this sense all of the terms in the
Skyrme energy density functional~\cite{engel1975} contribute to the measure.

An alternate meaning of the localization function relates to the kinetic energy and the Pauli
exclusion principle. In fact, the last two terms in Eq.~(\ref{eq:prob_D}) are the kinetic density for a complex valued single particle state of a
given spin $s$ and isospin $q$ that satisfies local gauge invariance~\cite{pittalis2010,li2020}
\begin{equation}
	\tau_{qs}^{\rm s.p.}= \frac{1}{4}\frac{\left|\boldsymbol{\nabla} \rho_{qs}\right|^2}{\rho_{qs}} + \frac{\left|\mathbf{j}_{qs}\right|^2}{\rho_{qs}}\,.
    \label{eq:tsp}
\end{equation}
The first term in Eq.~\eqref{eq:tsp} is related to the von Weizsacker kinetic-energy density~\cite{weizsacker1935}.
Equations~(\ref{eq:prob_D}) and~(\ref{eq:tsp}) give 
\begin{equation}
	D_{qs}=\tau_{qs}-\tau_{qs}^{\rm s.p.}\,,
\end{equation}
meaning that, for a single nucleon system, $D_{qs}=0$.
Consequently, $D_{qs}$ is a measure of the excess
of kinetic density due to the Pauli exclusion principle.
One can now define the Pauli kinetic energy (PKE) from the above expression as
\begin{equation}
	E_{qs}^\mathrm{P}=\frac{\hbar^2}{2m}\int d^3r\; D_{qs}(\mathbf{r})\,.\label{eq:EPauli}
\end{equation}
Note that this is the intrinsic PKE produced by all nucleons of any  nuclear system. 
So far, these equations are general and can be used to compute the PKE from different mean-field approaches by using the corresponding one-body density matrix.

For example, using a DCFHF one-body density matrix $\rho^{\mathrm{DCFHF}}(R)$ for nuclei at a distance $R$ in Eq.~(\ref{eq:prob_D}),
one can compute the total PKE with Eq.~(\ref{eq:EPauli}) from the resulting $D^{\mathrm{DCFHF}}_{qs}(R;\mathbf{r})$. 
However, the Pauli exclusion principle between nucleons belonging to the same nucleus do not contribute to the Pauli repulsion in the nucleus-nucleus potential. 
We then define the net PKE  for two nuclei
separated by a distance $R$ between their centers as
\begin{equation}\label{pke-1}
	\Delta E_{q\mu}^{\mathrm{P(F)}}(R)=\frac{\hbar^2}{2m}\sum_{s_\mu}\int d^3r\; \left[ D_{qs_{\mu}}^\mathrm{DCFHF}(\mathbf{r},R)- D_{qs_{\mu}}^\mathrm{FHF}(\mathbf{r},R)\right]\,,
\end{equation}
where we have subtracted the contribution of the PKE from the FHF approach. 
Indeed, the latter uses the same frozen density as DCFHF, but it neglects the Pauli exclusion principle between nucleons of different nuclei. 
We have also summed over the spin up and down components
for a given spin projection axis $\mu$.
The notation $\mathrm{P(F)}$ stands for ``Pauli in the Frozen approximation''.

The calculation of this PKE difference
with frozen densities allows one to identify the internuclear PKE without the presence of particle transfer and
other relaxation effects similar to methods used in
condensed matter physics and chemistry~\cite{wu2009,raupach2015,andrada2020}.
We note that $\Delta E^{\mathrm{P(F)}}(R)$ is zero
for large values of $R$ since the antisymmetrization of two well separated nuclei does not introduce
any change to local nuclear densities of Eq.~(\ref{eq:density_relation}) and thus both DCFHF and FHF give the intrinsic Pauli energy in this case.
We note also that, assuming that the PKE is independent of the choice of the spin quantization axis, summing over the isospin $q$ gives the total Pauli repulsion which should then be equivalent to the difference between the DCFHF and FHF potentials, $\sum_q \Delta E^{\mathrm{P(F)}}_{q\mu}(R)\simeq V_{\mathrm{DCFHF}}(R)-V_{\mathrm{FHF}}(R)$. 

We can similarly define the dynamical contribution to this net PKE by taking the difference between the DC-TDHF and DCFHF  localization measures
\begin{equation}\label{pke-D}
	\Delta E_{q\mu}^{\mathrm{P(D)}}(R)=\frac{\hbar^2}{2m}\sum_{s_\mu}\int d^3r\; \left[ D_{qs_{\mu}}^\mathrm{DC-TDHF}(\mathbf{r},R)- D_{qs_{\mu}}^\mathrm{DCFHF}(\mathbf{r},R)\right]\,,
\end{equation}
where $\mathrm{(D)}$ stands for ``dynamical contribution''. 
In this way we can easily compare the static (i.e., in the frozen approximation) and dynamical contributions to the Pauli repulsion. 

It is  interesting to visualize the NLF as it is also defined from the localization measure in Eq.~(\ref{eq:prob_D}). 
We first normalize the localization measure using~\cite{liu2019b,li2020}
\begin{equation}
	\mathpzc{D}_{qs_\mu}(\mathbf{r})=\frac{D_{qs_\mu}(\mathbf{r})}{\tau_{qs_\mu}^{\mathrm{TF}}(\mathbf{r})}\,, 
\end{equation}
where the normalization $\tau_{qs_\mu}^{\mathrm{TF}}(\mathbf{r})=\frac{3}{5}\left(6\pi^2\right)^{2/3}\rho_{qs_\mu}^{5/3}(\mathbf{r})$ is the Thomas-Fermi kinetic density.
The NLF can then be represented 
 either by $1/\mathpzc{D}_{qs_\mu}$ or by
\begin{equation}
	{C}_{qs_\mu}(\mathbf{r})=\left[1+\mathpzc{D}_{qs_\mu}^{2}\right]^{-1}\,.
	\label{eq:NLF}
\end{equation}
The advantage of the latter form is that it scales to be in the interval $[0,1]$, but otherwise both forms
show similar localization details. We also see that while this is called the NLF, it is  a visual
measure of the Pauli energy.

\subsection{Numerical details}

Calculations were done in a three-dimensional
Cartesian geometry with no symmetry assumptions using
the code of Ref.~\cite{umar2006c} and using the
Skyrme SLy4d interaction~\cite{kim1997}, which has been successful in 
describing various types of nuclear reactions~\cite{simenel2012,simenel2018}.
The three-dimensional Poisson equation for the Coulomb potential
is solved by using Fast-Fourier Transform techniques
and the Slater approximation is used for the Coulomb exchange term.
The static HF equations and the DCFHF
minimizations are implemented using the damped gradient
iteration method~\cite{bottcher1989}. The box size used for all the calculations
was chosen to be $60\times 30\times 30$~fm$^3$, with a mesh spacing of
$1.0$~fm in all directions. 
These values provide very accurate results due to the employment 
of sophisticated discretization techniques~\cite{umar1991a,umar1991b}.

\section{Results}\label{sec2}

\subsection{NLF and Pauli energy distribution \label{sec:NLF}}

\begin{figure}[!htb]
	\includegraphics*[width=8.6cm]{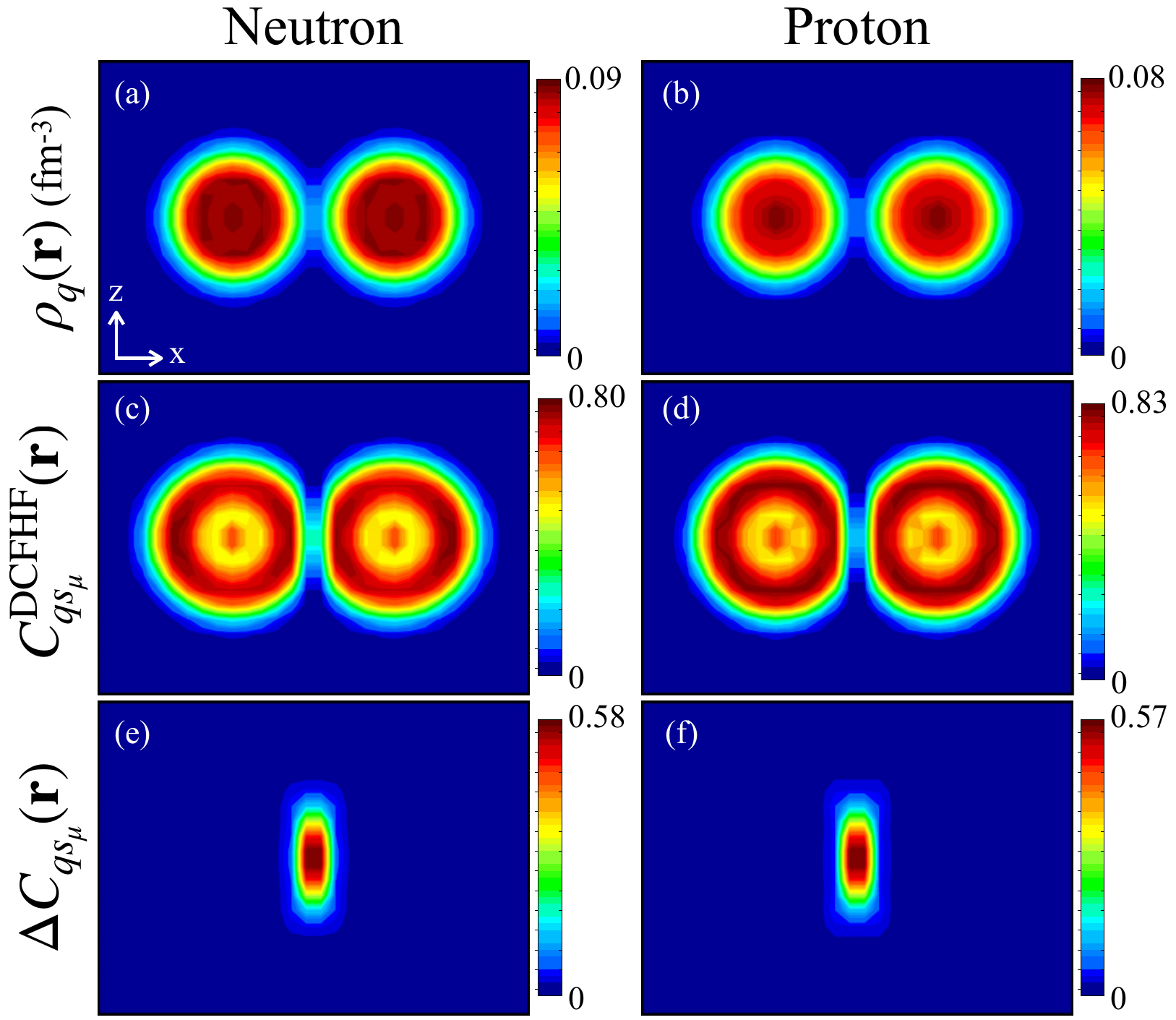}
	\caption{\protect Frozen neutron (a) and proton (b) HF densities in $^{48}$Ca$+^{48}$Ca for a distance  between the nuclei close to the barrier radius $R_B\simeq10.8$~fm.
	Only a sub-part of the numerical box is shown, with $29\times21$~fm$^2$ in the $(x,z)$ plane.
	Corresponding normalized NLF, computed from Eq.~\eqref{eq:NLF} for spin up (along the $\mu=z$ axis) with the DCFHF method, are plotted in (c) and (d). (The plots for spin down are similar). The difference $\Delta C_{qs_z}=C_{qs_z}^{\mathrm{FHF}}-C_{qs_z}^{\mathrm{DCFHF}}$ between FHF and DCFHF normalized NLF are plotted in (e) and (f). }
	\label{fig:nlfCa}
\end{figure}

\begin{figure}[!htb]
	\includegraphics*[width=8.6cm]{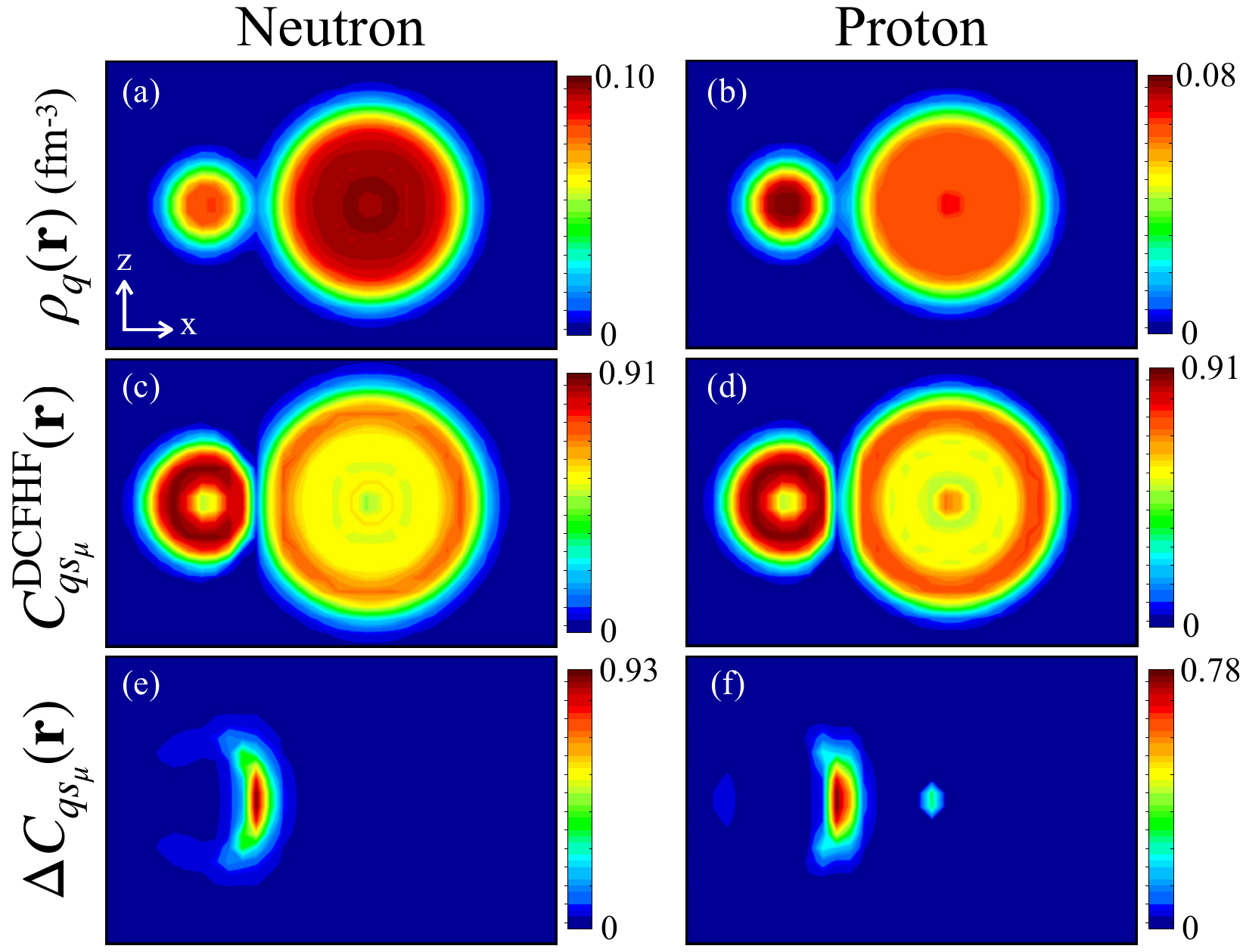}
	\caption{\protect Same as Fig.~\ref{fig:nlfCa} for $^{16}$O$+^{208}$Pb with $R_B\simeq 12$~fm. The part of the box that is represented covers $33\times21$~fm$^2$ in the $(x,z)$ plane.}
	\label{fig:nlfOPb}
\end{figure}

The outcome of a central heavy-ion collision at near barrier energy is 
highly sensitive to  the structure and dynamics of the system at the barrier radius $R_B$, i.e., when short-range nuclear attraction and long-range Coulomb repulsion compensate. 
The proton and neutron frozen densities in $^{48}$Ca$+^{48}$Ca and $^{16}$O$+^{208}$Pb are represented in panels (a) and (b) of Figs.~\ref{fig:nlfCa} and~\ref{fig:nlfOPb}, respectively. 
In both cases the neck density is small and does not exceed $20\%$ of the saturation density.
Note, however, that $R_B$ is determined from DC-TDHF potentials including shape polarization which may slightly increase the neck density.
Nevertheless, one would expect that the small neck density in the frozen calculations implies that the Pauli repulsion should remain small a priori. 
This expectation, however, is based on the assumption that the nucleons contributing to the neck density are essentially localized in the neck.

To verify this assumption, we plot the normalized NLF from DCFHF using Eq.~(\ref{eq:NLF}) in panels (c) and (d) of Figs.~\ref{fig:nlfCa} and~\ref{fig:nlfOPb}.
Interestingly, while the densities of the fragments add up in the neck region, the NLF behave differently with an apparent repulsion  flattening the facing surfaces of the NLF near contact. 
A possible interpretation is that the neck nucleons could be delocalized, i.e., belonging to both fragments. This would provide a microscopic support to the ``collectivization'' mechanism invoked by Zagrebaev~\cite{zagrebaev2001}.
However, this observation, in itself, does not necessarily imply a large Pauli repulsion. 
Nevertheless, it weakens the validity of the commonly used assumption that Pauli repulsion can be neglected when the density in the neck is small compared to saturation density. 

To get a deeper insight into the spatial distribution of the Pauli repulsion, we plot  the difference of the NLF computed from FHF and DCFHF in panels (e) and (f) of Figs.~\ref{fig:nlfCa} and~\ref{fig:nlfOPb}.
As expected, this difference is mostly found around the neck region as it is due to the symmetrization process which tends to delocalize the neck nucleons. 
As a result, the Pauli repulsion is mostly occurring in the neck region.
Note, however, that in the asymmetric $^{16}$O+$^{208}$Pb reaction, the spatial distribution of the Pauli repulsion also extends away from the neck. 
This distribution is also found be different for neutrons and protons. 
Quantitative comparisons between neutron and proton repulsion are discussed in the next section. 

\subsection{Pauli repulsion from potentials and NLF}

\begin{figure*}[!htb]
	\includegraphics*[width=18cm]{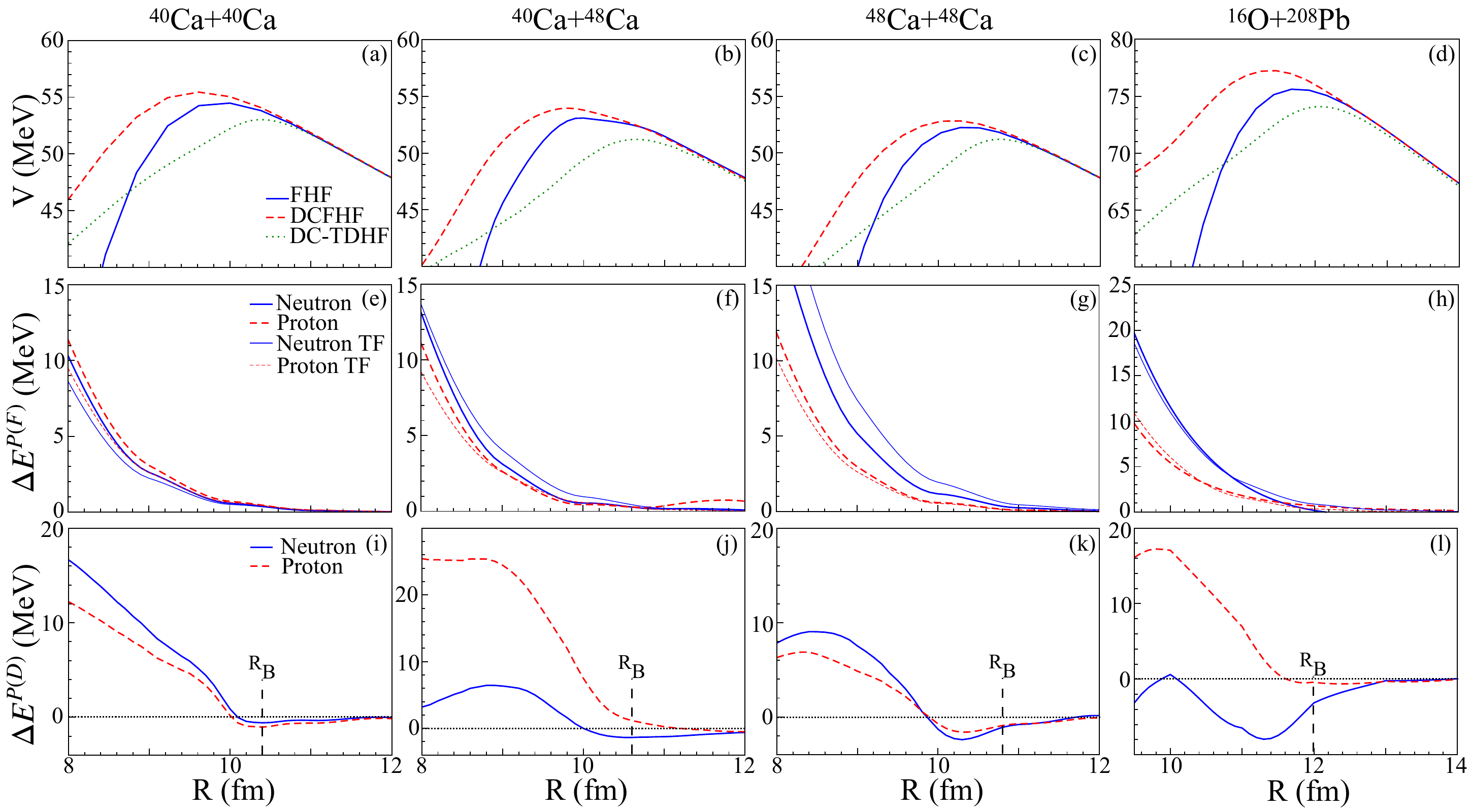}
	\caption{\protect (a-d) Nucleus-nucleus potentials in $^{40,48}$Ca$+^{40,48}$Ca and $^{16}$O$+^{208}$Pb computed from FHF, DCFHF, and DC-TDHF methods. (e-h) Neutron and proton contributions to the  Pauli repulsion from Eq.~(\ref{pke-1}) in the frozen approximation (thick lines). Replacing the DCFHF Pauli energy by the FHF one with Thomas-Fermi approximation of the kinetic energy in Eq.~(\ref{pke-1}) lead to results represented by thin lines. (i-l) Additional dynamical contributions to the Pauli repulsion computed from Eq.~(\ref{pke-D}). The line marked by $R_{B}$ indicates the (DC-TDHF) barrier peak for each system. The dotted horizontal lines show the location of zero PKE difference.} 
	\label{fig:Pauli}
\end{figure*}

Pauli repulsion is a contribution to the bare nucleus-nucleus potential which  becomes important at short distance when the nuclei overlap. 
The FHF and DCFHF potentials are represented  by solid and dashed lines, respectively, in Figs.~\ref{fig:Pauli}(a-d) for the $^{40,48}$Ca$+^{40,48}$Ca and $^{16}$O$+^{208}$Pb systems.
The Pauli repulsion included in DCFHF induces a widening of the barrier as well as a small increase of its height, up to $\sim1.6$~MeV in $^{16}$O$+^{208}$Pb.
To a large extent, this increase of the barrier height is compensated by dynamical polarization effects accounted for in the DC-TDHF potential [dotted lines in Fig.~\ref{fig:Pauli}(a-d)]. 
Indeed, coupling effects in these systems lower the average DC-TDHF barrier by few MeV. 
Nevertheless, the widening of the barrier due to Pauli repulsion is still present, indicating that it plays an important role in the inner barrier region. 

\begin{figure}[!htb]
	\includegraphics*[width=8.6cm]{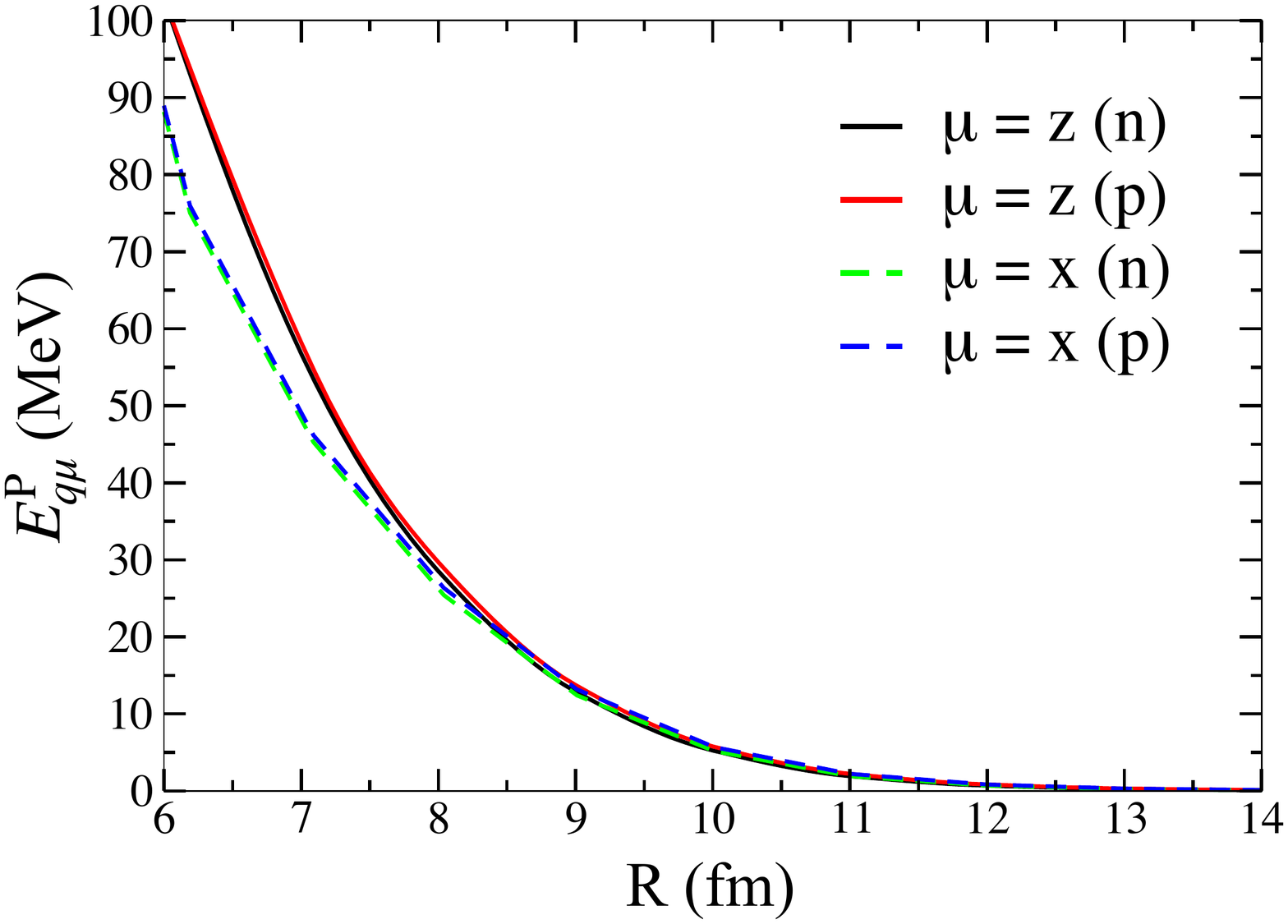}
	\caption{\protect Proton and neutron Pauli energy in $^{40}$Ca$+^{40}$Ca computed from DCFHF using Eq.~(\ref{eq:EPauli}) and summing over spin, $E^{\mathrm{P}}_{q\mu}=\sum_{s_\mu}E^{\mathrm{P}}_{qs_\mu}$. The solid and dashed lines are obtained with the quantization axis $\mu=z$ and $x$, respectively. }
	\label{fig:Polar}
\end{figure}

The NLF method allows to decompose the Pauli repulsion into proton and neutron contributions. 
This decomposition would make sense, however, only if the resulting Pauli energy does not strongly depend on the choice of the spin quantization axis $\mu$ in the region of interest. 
This dependence is studied in Fig.~\ref{fig:Polar} for $^{40}$Ca$+^{40}$Ca by comparing $\mu=z$ and $x$ spin quantization axes. 
Although the difference reaches $\sim10\%$ at very short distances, it remains small in the region studied in Fig.~\ref{fig:Pauli}, which is physically relevant for sub-barrier fusion reactions down to $\sim20\%$ below the barrier. 
We have chosen the spin projection axis to be
the $z-$axis (perpendicular to the collision axis $x$) in all the following calculations. In general, except at small distances (where the frozen density distributions become unphysical), we don't
see any significant dependence on the choice of the spin projection axis.

\begin{figure}[!htb]
	\includegraphics*[width=8.6cm]{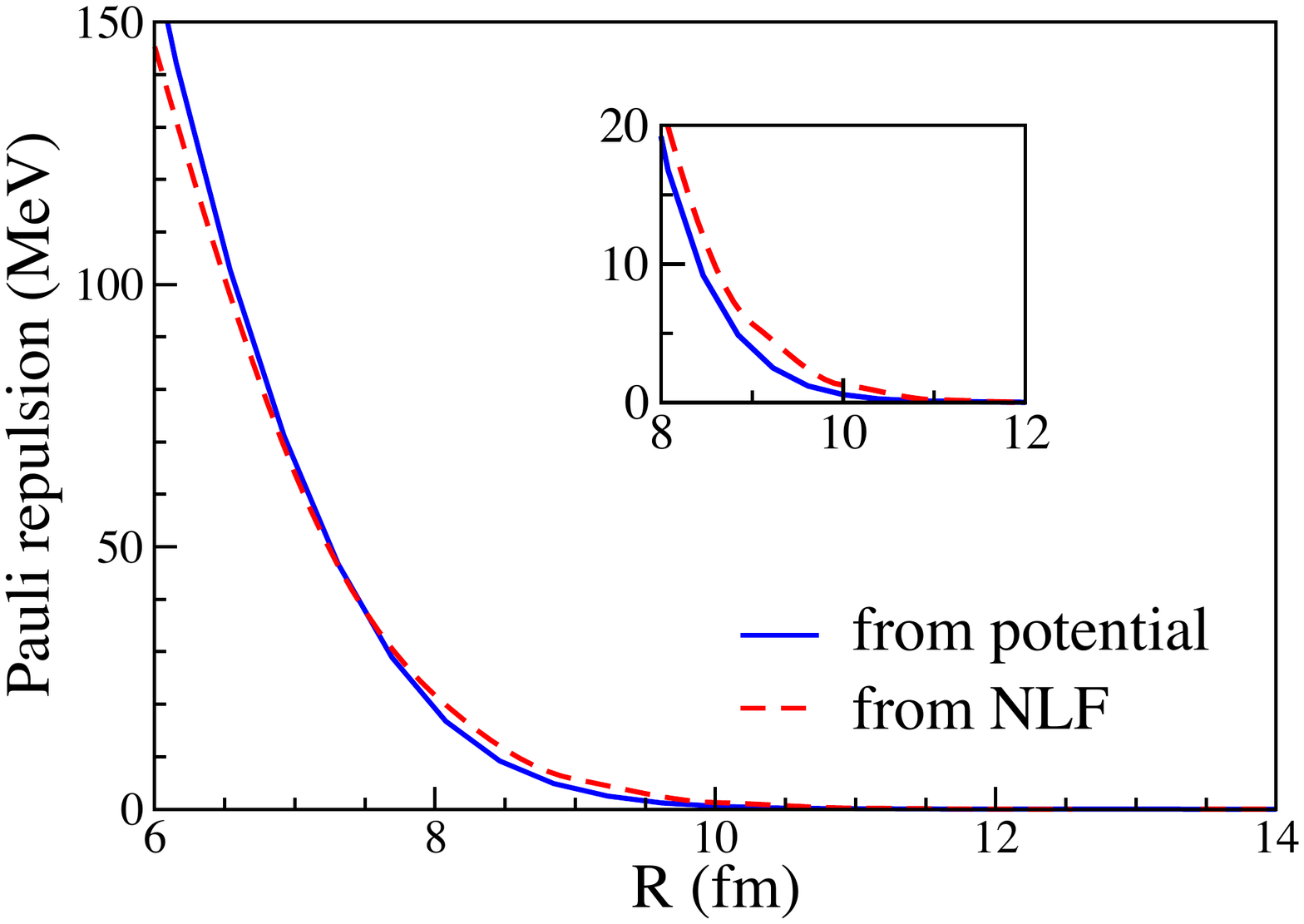}
	\caption{\protect Pauli repulsion in $^{40}$Ca$+^{40}$Ca. 
	The solid line is obtained from the potential difference $V^{\mathrm{DCFHF}}-V^{\mathrm{FHF}}$. The dashed line is the sum of neutron and proton net PKE from Eq.~(\ref{pke-1}) with $\mu=z$ spin quantization axis. The inset shows the same quantities in the barrier region. }
	\label{fig:V_NLF}
\end{figure}

Before we study in more details the proton and neutron contributions to the Pauli repulsion, we need to verify that the NLF method gives the expected Pauli repulsion. 
A comparison between Pauli repulsion calculations from the potential difference $V^{\mathrm{DCFHF}}-V^{\mathrm{FHF}}$ and using the NLF method in the frozen approximation [Eq.~(\ref{pke-1})] is shown in Fig.~\ref{fig:V_NLF} for $^{40}$Ca$+^{40}$Ca.
The overall agreement between both methods is good, although the NLF method slightly overestimates the Pauli repulsion in the barrier region (see inset of Fig.~\ref{fig:V_NLF}). 
It should be noted, however, that the localization measure in Eq.~(\ref{eq:prob_D}) is obtained within the local density approximation and that higher order terms in the expansion may account for this difference. 
Nevertheless, the calculated difference is small enough so that we can now turn our attention to investigating proton and neutron contributions to the Pauli repulsion with the NLF method. 

\subsection{Proton and neutron contributions to Pauli repulsion}

Proton and neutron contributions to the Pauli repulsion, computed from Eq.~(\ref{pke-1}) in the frozen approximation, are shown in Fig.~\ref{fig:Pauli}(e-h). 
All systems exhibit a sharp rise of Pauli repulsion inside the barrier. 
In all systems, the PKE acquires substantial values for small ion-ion separations (see also Fig.~\ref{fig:V_NLF} for $^{40}$Ca+$^{40}$Ca where the Pauli repulsion keeps increasing inside the barrier). This is the root of
the development of a potential pocket at short distances as the contribution from the PKE must overcome the sharp decline
of the frozen density FHF potentials. 

Both proton and neutron contributions remain very close in the $N=Z$ system $^{40}$Ca$+^{40}$Ca. 
Although the proton contribution  remains almost unchanged in all $^{40,48}$Ca$+^{40,48}$Ca systems, the Pauli repulsion significantly increases with the number of neutrons inside the barrier. 
For instance,  at $R\simeq9$~fm in $^{48}$Ca$+^{48}$Ca, which corresponds approximately to the inner turning point of a tunneling path at $\sim0.9V_B$ through the bare potential, the Pauli repulsion is still largely dominated by neutrons, with $\Delta E^{\mathrm{P(F)}}_{pz}(9\mbox{fm})\simeq 3$~MeV and  $\Delta E^{\mathrm{P(F)}}_{nz}(9\mbox{fm})\simeq 5$~MeV.
As a result, the ratio of the neutron contribution to the proton one largely exceeds the neutron-to-proton ratio. 
A similar behavior is observed in $^{16}$O$+^{208}$Pb where the neutron contribution is about twice that of protons inside the barrier. 
This can be interpreted as an effect of neutron skins which develop in neutron rich nuclei and which make the neutrons ``touch first'' near the barrier radius. 
The Pauli exclusion principle acts then first on neutrons in such systems. 
Consequently, one may expect a fusion hindrance due to the Pauli exclusion principle in neutron rich systems, in particular at sub-barrier energies. 

We finally note that, in the asymmetric systems $^{40}$Ca$+^{48}$Ca and $^{16}$O$+^{208}$Pb, the proton contribution exceeds the neutron one at large distance, i.e., at $R>R_B$.
The origin of this behavior is not clear. 
Note that it is not captured by the Thomas-Fermi approximation (see  Sec.~\ref{sec:TF}). 
We saw in Sec.~\ref{sec:NLF} that the Pauli exclusion principle tends to delocalize the neck nucleons. 
It is possible that this delocalization cost more energy for protons than neutrons, in particular in asymmetric systems.

\subsection{Test of the Thomas-Fermi approximation \label{sec:TF}}

It has been suggested by several groups~\cite{brink1975,brink1978,denisov2010} that, when constructing ion-ion potentials using frozen densities, the Pauli kinetic energy induced by antisymmetrization could be approximated by replacing the 
kinetic  density by the Thomas-Fermi expression 
$$\tau_{qs_\mu}^{\mathrm{TF}}(\mathbf{r})=\frac{3}{5}\left(6\pi^2\right)^{2/3}\rho_{qs_\mu}^{5/3}(\mathbf{r})\,.$$
We have tested this approach by replacing $D_{qs_{\mu}}^\mathrm{DCFHF}(\mathbf{r},R)$ in Eq.~(\ref{pke-1}) with the FHF expression evaluated with the Thomas-Fermi kinetic density. 
The results are plotted in Fig.~\ref{fig:Pauli}(e-h) with thin lines.  

Overall, the Thomas-Fermi approximation provides results in good agreement with the NLF method.
In particular, it underestimates the NLF calculation in $^{40}$Ca+$^{40}$Ca by less than 1~MeV (except at $R\sim8$~fm). 
The agreement is also good at short distance in $^{16}$O$+^{208}$Pb.
Near the barrier, however, the Thomas-Fermi approximation overestimates the neutron contribution by about 1 MeV in this system. 
As a result, it does not reproduce the dominant proton contribution at the barrier which is predicted by the NLF method in both $^{16}$O$+^{208}$Pb and $^{40}$Ca+$^{48}$Ca.
Another observation is that, in reactions with $^{48}$Ca, the splitting between neutron and proton contributions to the Pauli repulsion is somewhat overestimated by the Thomas-Fermi approximation. 

We conclude that, though the Thomas-Fermi approximation gives a reasonably good estimate of the Pauli repulsion, it does not always reproduce the various features of the proton and neutron contributions. 
In fact, in using the Thomas-Fermi approximation, one implicitly assumes that the Pauli repulsion only affects the kinetic density, while in fact it also have an impact on other terms such as the spin-orbit contribution to the energy~\cite{simenel2017}. 
It should also be noted that we only consider doubly-magic spherical nuclei which do not have pairing contributions at the mean-field level. 
Therefore, it would be interesting to test The Thomas-Fermi approximation  in mid-shell nuclei as well. 
This will be the purpose of a future work.

\subsection{Effect of dynamical rearrangement on Pauli energy}

As long as we are dealing with frozen densities the DCFHF method
embodies the aspects of antisymmmetrization to produce an optimum prediction of the PKE at the (static) mean-field level.
However, in a more realistic dynamical description of the reaction, the densities of the collision partners are no
longer frozen but start to interact with one another even before the barrier peak,
resulting in a  shape polarization and possible transfer of nucleons between the fragments.
Naturally, these processes
occur while also maintaining the Pauli principle. 
In the context of DFT, such dynamics are described by the TDHF (or TDDFT)  evolution. 
In the DC-TDHF methods, the nucleus-nucleus potential is determined from such time-dependent densities. 
A comparison between Pauli repulsion computed with  the frozen approximation and with DC-TDHF will then provide a deeper insight into the dynamical rearrangement of the Pauli kinetic energy.

Figures~\ref{fig:Pauli}(i-l) show the difference in Pauli energies calculated using DC-TDHF and DCFHF
methods as a function of the internuclear distance $R$.
In  all systems, the total Pauli repulsion (summed over proton and neutron contributions) is reduced near the barrier radius $R_B$  once dynamics is included. 
As TDHF equations are obtained from a stationary action principle (usually leading to a minimum action),
and because the PKE increases the action, it is natural for the system to explore paths in which the density is rearranged to  reduce the PKE.
At shorter distances, however, an increase of the total PKE is observed. 
It should be noted that the density distributions obtained at small $R$ are usually very different to the frozen densities, and that the difference of PKE at $R<R_B$ may not be always meaningful. 
Nevertheless, the observed increase of Pauli repulsion due to dynamics at short distance could also have a physical origin, such as a dynamical collectivization of the nucleons (increasing the PKE) when the collision partners merge into a single system. 

We now turn our attention to the difference between proton and neutron contributions to the dynamical rearrangement of the net PKE in Figs.~\ref{fig:Pauli}(i-l). 
Both species behave similarly in the symmetric $^{40}$Ca$+^{40}$Ca and $^{48}$Ca$+^{48}$Ca reactions, though in both cases neutrons induce a slightly stronger repulsive dynamical contribution at short distance. 
In the asymmetric $^{40}$Ca$+^{48}$Ca and $^{16}$O$+^{208}$Pb systems, however, the dynamics induce a strong net PKE for the protons, while the neutron contribution remains small, and even attractive in the $^{16}$O$+^{208}$Pb system. 
In terms of dynamics, the main difference between the symmetric and asymmetric collisions is the opening of transfer channels in the latter, driven by an $N/Z$ equilibration between the nuclei. This is a rapid process, with a timescale of the order of $1$~zs~\cite{simenel2020}, that occurs as soon as the nuclei are in contact. 
In both reactions, we expect a (multi)proton transfer from the light ($N=Z$) to the heavy ($N>Z$) nucleus, while neutrons are expected to flow the other way, as demonstrated in experimental studies of $^{16}$O+$^{208}$Pb~\cite{evers2011,rafferty2016}.
In both reactions, the Pauli exclusion principle prevents protons to be transferred to similar states as they are already occupied. 
This could be a possible explanation for the increase of the net PKE due to proton dynamics. 
On the other hand, neutrons can be transferred to unoccupied final states with similar structure to the  initial state. 
In this case, the Pauli exclusion principle is not ``opposed'' to the transfer and no particular rise of the PKE is expected. 
This simple interpretation could be tested with TDHF through a microscopic analysis of the initial and final states involved in the transfer  (see, e.g., Fig.~22 of~\cite{simenel2018}).
Such tests will be the subject of future work.

\section{Conclusions}\label{sec3}
The impact of the Pauli exclusion principle on various models and approaches of calculating the
interaction of two nuclei has been a topic of interest for a long time. This is mainly predicated
by the fact that, in most approaches, the influence of antisymmetrization has been an after thought.
Furthermore it has been argued that such effects are minimal at the barrier peak. Naturally, this was first
found to be not true in alpha-nucleus potentials and subsequently at higher bombarding
energies or for deep sub-barrier energies. Various remedies included the simple antisymmetrization of
states between the two nuclei, which was found to require a normalization factor since the resulting
states do not necessarily represent the lowest energy configuration, or the phenomenological shallow
potential approach that introduced empirically adjusted potential pockets~\cite{misicu2006}.

The recently proposed DCFHF method has been employed to compute bare nucleus-nucleus potentials in $^{40,48}$Ca$+^{40,48}$Ca and $^{16}$O$+^{208}$Pb reactions. 
This method not only does the necessary
antisymmmetrization but also minimizes the energy of the system by keeping the density
frozen but adjusting the underlying orbitals.
In addition to account for the Pauli exclusion principle exactly (at the mean-field level), an appealing aspect of this method is that it does not require additional parameters other than those of the Skyrme EDF  used to build the nuclear mean-field. 
By comparing with the more standard FHF (which neglects the Pauli exclusion principle between nucleons of different nuclei), we can compute the Pauli repulsion between the nuclei, which effects is essentially to widen the barrier and increase its height, thus hindering sub-barrier fusion.  

We have employed the Pauli kinetic energy expression that arises from the study
of the nuclear localization function within the realm of the density functional theory.
The spatial distribution of the Pauli kinetic energy shows that the repulsion occurs essentially in the neck between the fragments when the latter are at a distance equivalent to the barrier radius. The Pauli kinetic energy has also been used to decompose the Pauli repulsion into its proton and neutron contributions, 
We have shown that, inside the barrier, the contribution due to neutron dominates in neutron-rich systems, which is expected to further hinder sub-barrier fusion in such systems.

Additional effects due to the dynamics of the Pauli kinetic energy have been studied with the DC-TDHF method. 
We showed that, near the barrier, the system is usually able to find a path that reduces  the Pauli repulsion. 
Inside the barrier, however, the net effect of the dynamics is essentially to increase the Pauli kinetic energy when the nuclei merge and the nucleons are subject to a collectivization process.
A strong difference between proton and neutron dynamical contributions to the PKE is observed in the asymmetric systems we have studied. 
It is interpreted as an effect of multinucleon transfer induced by a rapid $N/Z$ equilibration.
Note that the availability of several states following transfer with positive $Q$-values could counterbalance the Pauli repulsion
that, in general, is predicted to reduce tunneling probability inside the Coulomb barrier~\cite{stefanini2017,stefanini2019}.

Future directions include the study of Pauli repulsion in mid-shell nuclei to investigate the effects of pairing and deformation. 
In addition, it would be interesting to get a deeper insight into the interplay between transfer and Pauli kinetic energy through microscopic analysis of single-particle states involved in the transfer mechanism.

\begin{acknowledgments}
We thank W. Nazarewicz for useful discussions.
This work has been supported by the U.S. Department of Energy under grant No.
DE-SC0013847 with Vanderbilt University
and by the NNSA Cooperative Agreement DE-NA0003841, and by the
Australian Research Councils Grant No. DP190100256.
\end{acknowledgments}

\bibliography{VU_bibtex_master.bib}


\end{document}